\documentclass[preprint2]{aastex}
\usepackage[twocolumn, textwidth=18cm, columnsep=.75cm, margin=2.0cm]{geometry}
\usepackage{graphicx}
\usepackage{amsmath}
\usepackage{stfloats} 

\begin{document}

\title{Attaining Doppler Precision of 10 cm $\rm s^{-1}$ with a Lock-In Amplified Spectrometer}

\author{Rebecca Jensen-Clem\altaffilmark{1}, Philip S. Muirhead\altaffilmark{2}, Michael Bottom\altaffilmark{1}, J. Kent Wallace\altaffilmark{3}, Gautam Vasisht\altaffilmark{3}, John Asher Johnson\altaffilmark{4}}

\altaffiltext{1}{\footnotesize{California Institute of Technology, Pasadena, CA, USA 91101.}}
\altaffiltext{2}{Department of Astronomy, Boston University, Boston, MA 02215.}
\altaffiltext{3}{Jet Propulsion Laboratory, California Institute of Technology, 4800 Oak Grove Drive, Pasadena, CA, USA 91109.}
\altaffiltext{4}{Harvard-Smithsonian Center for Astrophysics, 60 Garden St., Cambridge, MA, USA 02138}

\begin{abstract}
We explore the radial velocity performance benefits of coupling starlight to a fast-scanning interferometer and a fast-readout spectrometer with zero readout noise.  By rapidly scanning an interferometer we can decouple wavelength calibration errors from precise radial velocity measurements, exploiting the advantages of lock-in amplification. In a Bayesian framework, we investigate the correlation between wavelength calibration errors and resulting radial velocity errors. We construct an end-to-end simulation of this approach to address the feasibility of achieving 10 cm $\rm s^{-1}$ radial velocity precision on a typical Sun-like star using existing, 5-meter-class telescopes.  We find that such a precision can be reached in a single night, opening up possibilities for ground-based detections of Earth-Sun analog systems. 
\vspace{0.5cm}
\end{abstract}

\section{Introduction}
\label{sec:intro}

Recent results from NASA's \emph{Kepler} mission indicate that Sun-like stars are teeming with rocky exoplanets \citep[][]{Koch2010, Borucki2011, howard_planet_2012, Batalha2013, Petigura2013}. According to statistics derived from \emph{Kepler's} discoveries, and consistent with prior exoplanet searches, ground-based radial velocity and transit surveys targeting nearby stars are approaching the performance edge of discovering hundreds of exoplanets suitable for detailed atmospheric studies \citep[e.g.][]{Deming2009, Howard2010, dressing_occurrence_2013,berta_constraints_2013}.  NASA's future Transiting Exoplanet Survey Satellite (TESS) will likely find those rocky planets that transit nearby stars in short-period orbits \citep{ricker_transiting_2014}.  TESS will determine the radii of exoplanets transiting nearby stars.  With precise radial velocity measurements of the TESS discoveries, the exoplanet masses, mean densities and surface gravities can be determined, thereby constraining their interior structures and atmospheric characteristics.

The future of exoplanet science is therefore promising; however, the precision needed to detect nearby rocky planets with radial velocities (RVs), and measure the masses of those found to transit with TESS, is daunting.  For example, the Earth introduces an 8.9 cm $\rm s^{-1}$ semi-amplitude Doppler reflex motion on the Sun. To date, the smallest semi-amplitude radial velocity measured on a star is 51 cm $\rm s^{-1}$ on $\alpha$ Centauri Bb, indicating the presence of an exoplanet with a minimum mass of 1.13 M$_{\oplus}$, and a semi-major axis of 0.04 AU  ~\citep{dumusque_earth-mass_2012}.  The planet's discovery has yet to be confirmed by independent instruments or techniques, so $\alpha$ Centauri Bb is widely regarded as an exoplanet candidate. Nevertheless, the measurement indicates the state-of-the-art in stellar radial velocity precision. The detection of $\alpha$ Centauri Bb requires both astrophysical and instrumental noise corrections. Known astrophysical noise sources such as stellar oscillation modes, granulation, and activity signals can be individually removed through careful choices of exposure times and observing cadence, as well as modeling each effect separately \citep[][]{Dumusque2011}. ~\citet{pepe_harps_2011} describe how instrumental noise sources in  HARPS, such as changes to temperature, pressure, and illumination, are addressed through high cadence, long time baseline observations of nearby, slowly rotating stars with little known activity. To this end, HARPS observed HD 85512 for 7.5 years, obtaining 185 RV data points and discovering a 3.5 M$_{\oplus}$ planet. The standard deviation of the data residuals after subtracting the model of the planet was measured to be 0.75 m $\rm s^{-1}$. While this value could encompass uncharacterized astrophysical noises, 0.75 m $\rm s^{-1}$ is taken to be representative of the instrumental noise in HARPS. In order to reduce the state-of-the-art RV semi-amplitude from 51 cm $\rm s^{-1}$ ($\alpha$ Centauri Bb) to 8.9 cm $\rm s^{-1}$ (Earth-Sun analog), these instrumental noise sources must be understood and eliminated. 

Eliminating the effects of instrumental noise on radial velocity measurements requires ever more precise wavelength calibration techniques. Historically, iodine absorption cells and thorium-argon (ThAr) lamps have been the principal wavelength calibration tools. Starlight passing through an iodine gas cell results in a dense series of iodine absorption lines superimposed onto the stellar spectrum. This method has been implemented on HIRES, where planet semi-amplitudes as small as 1.89 m $\rm s^{-1}$ have been detected ~\citep[HD 156668b,][]{howard_nasa-uc_2011}. The small number of absorption lines at red and near-IR wavelengths, however, limits the use of iodine cell calibration for redder stars. Furthermore, the technique requires complex modeling of the combined iodine/stellar spectrum, which in turn requires high signal to noise data ~\citep{lovis_new_2007}. As an alternative to the iodine cell approach, fiber-fed spectrographs have used ThAr lamps as a simultaneous wavelength reference ~\citep{baranne_elodie:_1996}. Starlight and ThAr lamp light are fed into the spectrograph through separate fibers such that their spectra are simultaneously recorded onto the detector. While ThAr provides more lines in the near-IR than iodine, the non-uniformity in line spacing and long-term variability of ThAr sources limits the technique. 

Recently, laser frequency combs (LFCs) have provided large numbers of equally spaced, stable lines over a wide range of wavelengths ~\citep[e.g.][]{murphy_high-precision_2007, Steinmetz2008}. LFCs fix the phase of standing waves inside a laser cavity such that the waves periodically interfere constructively, producing bursts of light. The time  between bursts can be accurately controlled using an atomic clock. In the frequency domain, the laser therefore produces a series of equally spaced spectral lines. 

An experimental LFC installed on HARPS in 2010 has yielded unprecedented RV stability over short timescales ~\citep{wilken_spectrograph_2012}. The HARPS spectrograph is fed by two multimode fibers, or channels; one channel is coupled to starlight, and the other to light from the LFC. In ~\cite{wilken_spectrograph_2012}, the HARPS team fed LFC light through both fibers, and differenced the two channels to measure the instrumental drift. By optimally binning 20-30s exposures for 4 minutes, the limiting RV precision due to instabilities between the two channels was found to be 2.5 cm $\rm s^{-1}$. Between November of 2010 and January of 2011, however, synchronous drift in the channels induced a standard deviation of 34 cm $\rm s^{-1}$ between data taken in the two time periods. While planned changes to the environmental conditions surrounding HARPS made this 34 cm $\rm s^{-1}$ drift measurement a slight underestimate of the long term stability, the two channels nevertheless experienced long term drifts larger than the 2.5 cm $\rm s^{-1}$ short term limiting RV precision. These drifts are thought to be due to uncontrolled, long timescale instrumental changes. 

An innovative approach to acquiring Doppler measurements of stars was proposed by \citet{Erskine2003}. By introducing a Michelson or Mach-Zender interferometer into the optical path before a traditional spectrometer, \citet{Erskine2003} suggested high Doppler precision could be achieved with instrumental fluctuations minimized. The technique, called externally dispersed interferometry (EDI), requires that the optical path difference in the interferometer be modulated, either spatially or temporally. He suggested both a spatial modulation, by tilting one of the interferometer cavity mirrors, and a temporal modulation, by physically moving one of the cavity mirrors.  The phases of the resulting fringe patterns, whether temporal or spatial, are highly sensitive to changes in the Doppler velocity of the target star.  As noted by \citet{Erskine2003}, the modulation of a spectrum into a fringe pattern provides a robust way of eliminating systematic effects, such as fixed-pattern noise, in the resulting radial velocity measurements.

Since then, multiple groups have implemented EDI systems.  \citet{Ge2006} and \citet{mahadevan_measuring_2008} describe an instrument called the Exoplanet Tracker (ET) which uses a tilted cavity mirror without temporal modulation.  \citet{Vaneyken2010} describe the theory and implementation of ET and a similar instrument called the Multi-object APO Radial Velocity Exoplanet Large-area Survey (MARVELS).  \citet{Hajian2007} describe the dispersed Fourier Transform Spectrometer (dFTS), designed to acquire high-resolution spectra rather than measure precise radial velocities, also achieved by \citet{Erskine2003b}.  And finally, \citet{muirhead_precise_2011} describe the TripleSpec Exoplanet Discovery Instrument (TEDI), which used temporal rather than spatial modulation to map out the fringes.

A particularly interesting result from \citet{muirhead_precise_2011} was the achievement of 30 m $\rm s^{-1}$ of RMS radial velocity performance without any calibration of the spectrometer point-spread function.  Typically, precise radial velocity spectrometers involve a significant amount of effort to stabilize the spectrometer PSF because uncalibrated asymmetries in the PSF are degenerate with wavelength calibration errors, and result in spurious radial velocity measurements on the target star.  \citet{muirhead_precise_2011} show that the reduced importance of wavelength calibration is a direct result of the radial velocity measurement being encoded in the phase of a sinusoid varying within each pixel, rather than the change in flux across several pixels, as is the case for traditional spectroscopy as well as for ET and MARVELS, which use spatial modulation across pixels rather than temporal modulation within a pixel.

However, a significant challenge to temporal modulation techniques involves effects from readout noise.  The robustness against errors in wavelength calibration is due to modulation taking place on short timescales, and the PSF fluctuations taking place over longer time scales.  PSF fluctuations or wavelength calibration errors that occur over the same timescale as the modulation are not removed.  Fast scanning is therefore desired.  Readout noise in large format detectors limits the speed at which one can scan and readout the detector in the spectrometer.  In \citet{muirhead_precise_2011}, the authors took 30 second exposures to ensure that the dominant noise source was photon noise, rather than detector readout noise.

Recent developments in large format, high frame rate and low-readout noise detectors motivate studies of temporal modulation at very high frame rates.  Electron multiplying charge-coupled devices (EMCCDs) and complementary metal-oxide-semiconductor (CMOS) detectors can be manufactured and operated to have inherently low readout noise ($<$ 1 electron per pixel) and the ability to expose and readout at very high frame rates (~100 megapixels per second).  With a high enough gain enabled, EMCCDs have effectively zero readout noise.  These detectors are becoming commonplace in astronomical instruments \citep[e.g.][]{Dhillon2001,baranec_rise_2013}, motivating a study of the advantages of fast-modulation EDI for ultra-precise Doppler velocimetry.

In this paper, we show that the act of modulating a spectrum in time effectively decouples wavelength calibration errors from radial velocity measurements. Therefore, the nature of the radial velocity measurement is fundamentally different than that obtained from conventional spectroscopy, where wavelength calibration completely determines the resulting radial velocity measurement. Instead, the radial velocity precision relies upon precise knowledge of the interferometer positions\footnote{A possible implementation could involve a combination of fine pathlength control with precision PZTs and fine sensing using a co-propagating laser with exceptional wavelength stability.}. We augment previous theoretical studies of the EDI technique \citep[e.g.][]{Erskine2003, Vaneyken2010} by specifically introducing and recovering wavelength calibration errors into simulated data. We also investigate the effects of interferometer position errors on radial velocity precision. Finally, we simulate the performance of a time-varying interferometric system on a medium-sized telescope to determine the feasibility of an Earth-Sun analog detection. 

\section{Theory}
\label{sec:theory}

Following the discussion in \citet{Erskine2003} and also \citet{muirhead_precise_2011}, we derive the relationship between a temporally modulated spectrum and a radial velocity measurement. We use this relationship to demonstrate the decoupling of wavelength calibration errors and RV measurements. 

The intensity of the stellar spectrum recorded by the detector depends on both the wavelength and the interferometer delay. The interferometer delay is assumed to vary sinusoidally with time, therefore acting as a sinusoidal transmission comb. For a wave number $\nu$, and delay $\tau$, the measured intensity is given by
\begin{equation}
I_{\nu, \tau}=[S_{\nu}(1+\cos{(2 \pi \tau \nu)})]*R_{\nu}
\label{eqn:intensity}
\end{equation}
where $S_\nu$ is the intrinsic stellar spectrum, $(1+\cos{(2 \pi \tau \nu)})$ represents the transmission comb, $R_{\nu}$ is the spectrograph line spread function, and $*$ represents the convolution. While most practical interferometer configurations will result in two output beams, we assume here that the beams have been combined (for example, by placing a spectrograph and detector at each output and combining the intensities in post-processing). 

The interferometer delay $\tau$ can be expressed as the sum of a constant bulk delay, $\tau_0$, and a smaller time-varying phase shift $\Delta \tau$. Taking $\tau = \tau_0 + \Delta \tau$, and applying trigonometric identities, the measured intensity can be re-written as:
\begin{equation}
I_{\nu, \tau_0, \Delta \tau} = A_{\nu}+B_{\nu}\cos{(2 \pi \Delta \tau \nu)} - C_{\nu}\sin{(2 \pi \Delta \tau \nu)}.
\label{eqn:decomposed_intensity}
\end{equation}
The coefficients $A_{\nu}$, $B_{\nu}$, and $C_{\nu}$ are defined as 
\begin{equation}
A_{\nu} = S_{\nu}*R_{\nu},
\label{eqn:A}
\end{equation}
\begin{equation}
B_{\nu} = [S_{\nu}cos(2 \pi \tau_0 \nu)]*R_{\nu},
\label{eqn:B}
\end{equation}
\begin{equation}
C_{\nu} = [S_{\nu}\sin{(2 \pi \tau_0 \nu)}]*R_{\nu},
\label{eqn:C}
\end{equation}
where $A_{\nu}$ is the non-modulated spectrum, while $B_{\nu}$ and $C_{\nu}$ describe the interference between the stellar spectrum and the transmission comb. These coefficients can be determined by varying the interferometer delay $\Delta \tau$, and observing the intensity $I_{\nu, \tau_0, \Delta \tau}$. A data set, or ``scan" is a list of intensities at a range of $\Delta \tau$ and $\nu$ values.  

In order to compare the fitted coefficients between two scans, $B_{\nu}$ and $C_{\nu}$ are combined into a ``complex visibility:"
\begin{equation}
B_{\nu} - i C_{\nu} = [S_{\nu}e^{-i 2 \pi \tau_0 \nu}]*R_{\nu}
\label{eqn:comvis}
\end{equation}
The complex visibility is used to compare two scans, where one has been Doppler shifted by $\Delta \nu = (\Delta RV /c) \nu$. $\Delta \nu$ and $1/\tau_0$ are assumed to be small compared to a resolution element of the spectrograph. By applying the Fourier convolution and shift theorems, the Doppler shifted ``epoch" complex visibility $B_{\nu}^{1} - i C_{\nu}^{1}$ and the unshifted ``template" visibility $B_{\nu}^{0} - i C_{\nu}^{0}$ are related by an exponential phase:
\begin{equation}
 \label{eqn:compare}
\begin{split}
B_{\nu}^{1} - i C_{\nu}^{1} & = [B_{\nu}^{0} - i C_{\nu}^{0}]e^{-i 2 \pi \tau_0 \Delta \nu}\\ 
& =  [B_{\nu}^{0} - i C_{\nu}^{0}]e^{-i 2 \pi \tau_0 (\Delta RV)\nu/c}.
\end{split}
\end{equation}


For clarity, the discussion above applies to small radial velocity shifts only, due to the use of the Fourier shift theorem. On-sky observations, however, will be affected by the Earth's motion relative to the barycenter of the solar system, producing RV shifts on the order of 10 km $\rm s^{-1}$. \citet{muirhead_precise_2011} describe the detailed derivation of the relationship between the epoch and template complex visibilities, arriving at the following generalization of Equation \ref{eqn:compare}:

\begin{equation}
B_{\nu}^{1} - i C_{\nu}^{1} = e^{i 2 \pi \tau_0 \Delta \nu} \left[ (B_{\nu}^{0} - i C_{\nu}^{0} )e^{-i 2 \pi \tau_0 \Delta \nu} \right ]_{\nu \rightarrow \nu + \Delta \nu}
\label{eqn:compare_general}
\end{equation}

Therefore, a Doppler shift changes the phase of the complex visibility (for example, an Earth-Sun analog would result in a phase shift of $3.4\times 10^{-5}$ radians). A wavelength calibration error re-assigns the phase values to a different wavelength grid without changing the value of the phase itself. The radial velocity measurements are therefore decoupled from wavelength calibration changes; instead, the RV precision relies upon precise knowledge of the interferometer positions, described by $\Delta \tau$ and $\tau_{0}$. Section \ref{sec:sim_noise} describes the effects of wavelength calibration and interferometer position errors in detail.

\section{Simulation Architecture}
\label{sec:sim}

\begin{figure}[b]
  \centering
    \includegraphics[width=0.5\textwidth]{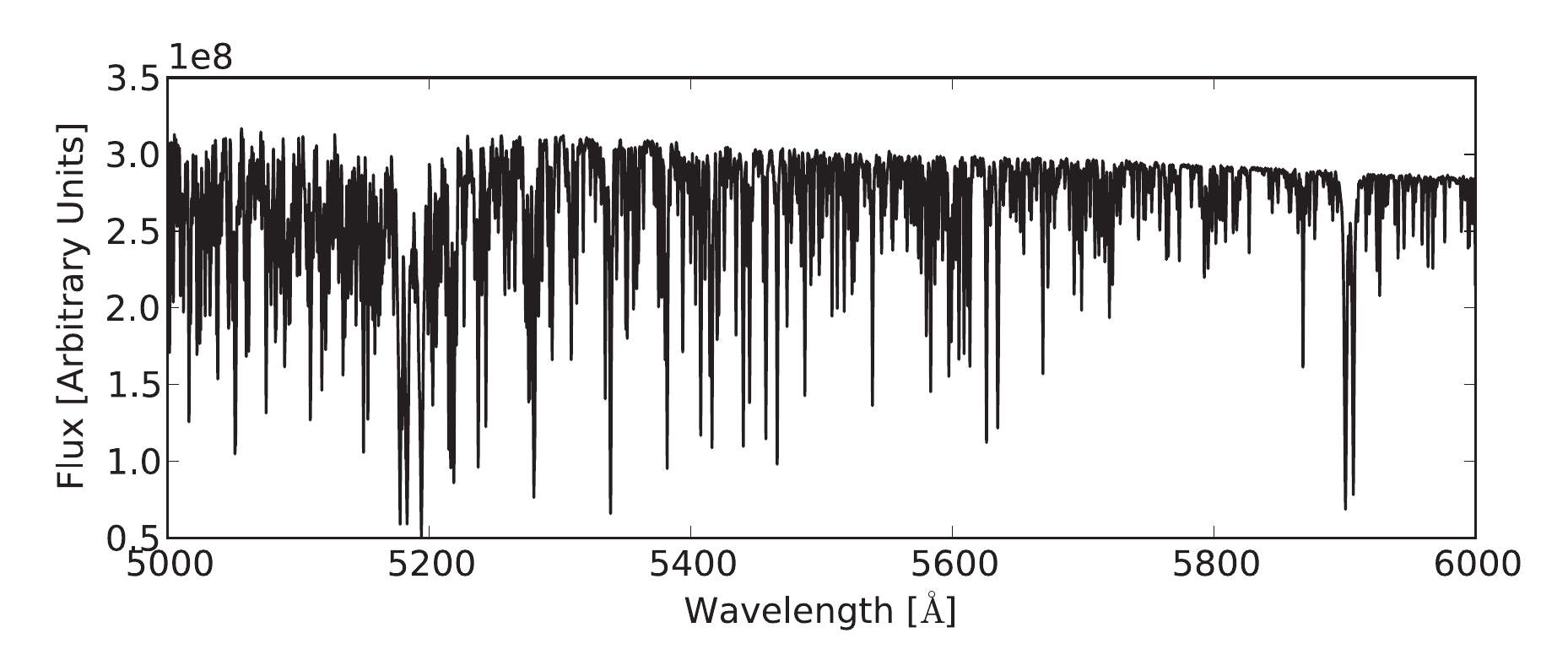}
    \caption{\small A Horizontal cut through Figure \ref{fig:smile} reveals the simulated stellar spectrum. Note that for ease of viewing, Figure \ref{fig:smile} shows the mean subtracted fringes.}
    \label{fig:smile_cut}
\end{figure}

In order to quantify instrumental contributions to radial velocity precision, we construct an end-to-end simulation of the radial velocity reconstruction process described in Section \ref{sec:theory}.The flux entering the instrument is modeled using a simulated solar spectrum ~\citep{coelho_spectral_2007}. We multiply the stellar spectrum by the interferometer comb, which is modeled as a sinusoidal transmission comb as a function of wavenumber $\nu$ and interferometer delay $\tau$ (see Equation \ref{eqn:intensity}). The result is therefore a series of spectra multiplied by transmission combs of different interferometer delays. 

\begin{figure}[t]
    \includegraphics[width=0.5\textwidth]{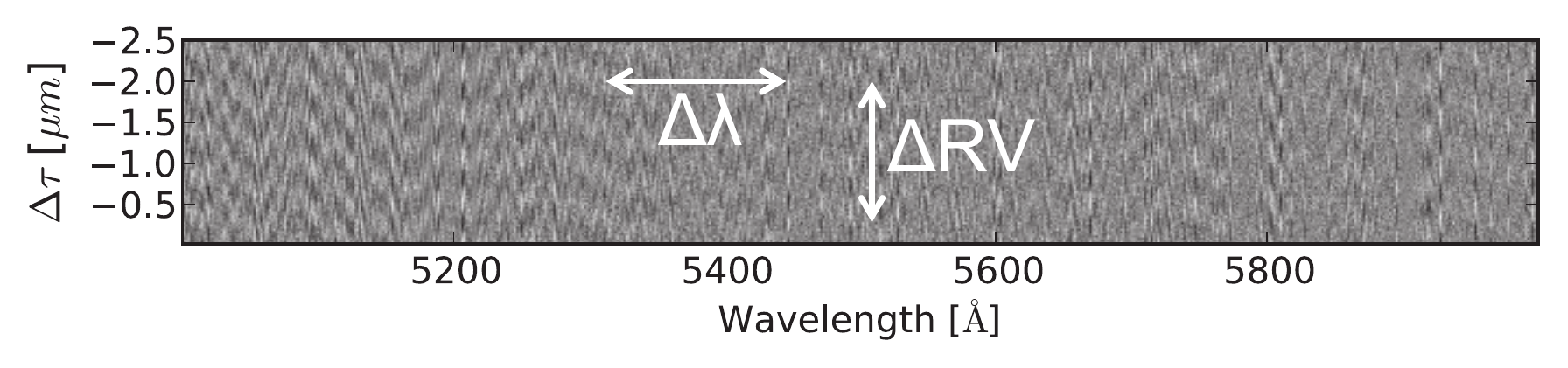}
    \caption{\small A simulated mean subtracted spectrum vs. wavelength and $\Delta \tau$ for SNR = 500. The x-axis corresponds wavelength, or pixels on the spectrometer detector.  The y-axis corresponds to changes in the optical path difference of the interferometer, and therefore represents time as the interferometer optical path difference is modulated temporally.  The regions of high fringe contrast correspond to absorption lines in the stellar spectrum.  An error in the wavelength calibration is fundamentally different from a change in the radial velocity of the target star, as long as the wavelength calibration error does not occur during an interferometer scan.}
    \label{fig:smile}
\end{figure}

The product of the stellar spectrum and interferometer transmission comb is convolved with the spectrograph instrument profile (IP). The spectrograph IP is modeled as a normal distribution with a width corresponding to a spectrograph resolution of R=10000. The resolution was chosen to be low in order to avoid resolving the interferometer comb; in order for the phase of the interference fringes to correspond to a radial velocity change, the comb must remain unresolved. The sampling is assumed to be 4 pixels per resolution element. The result is a stellar spectrum recorded at each interferometer step (Figure \ref{fig:smile_cut}). The final data product can be represented by a three dimensional map of interferometer delay, wavelength, and observed intensity (Figure \ref{fig:smile}). 

It is clear from the axes of Figure \ref{fig:smile} that wavelength calibration and radial velocity information are separated; the y-axis represents time (the time-varying position of the interferometer), while the x-axis represents detector pixels (the wavelength information introduced by the spectrograph). A radial velocity shift will therefore result in a vertical shift as the phase of the interference fringes changes, whereas a change in wavelength calibration will result in a global horizontal shift as the spectrum's wavelength is re-assigned. 

Each vertical cut through Figure \ref{fig:smile} represents the intensity as a function of phase delay at a given wavenumber described by Equation \ref{eqn:decomposed_intensity}. Noise is added to the intensity at each phase delay by adding values drawn from a Gaussian distribution with a mean of zero and a standard deviation given by the spectrum's mean divided by the desired signal to noise ratio. For clarity, we refer to this as Poisson noise, but our simplifications (Gaussian statistics with a constant standard deviation) should be noted. The $A_\nu$, $B_\nu$, and $C_\nu$ coefficients are then fit to each vertical cut, defining the complex visibilities (Equation \ref{eqn:comvis}). 

\begin{figure}[t]
  \centering
      \includegraphics[width=0.5\textwidth]{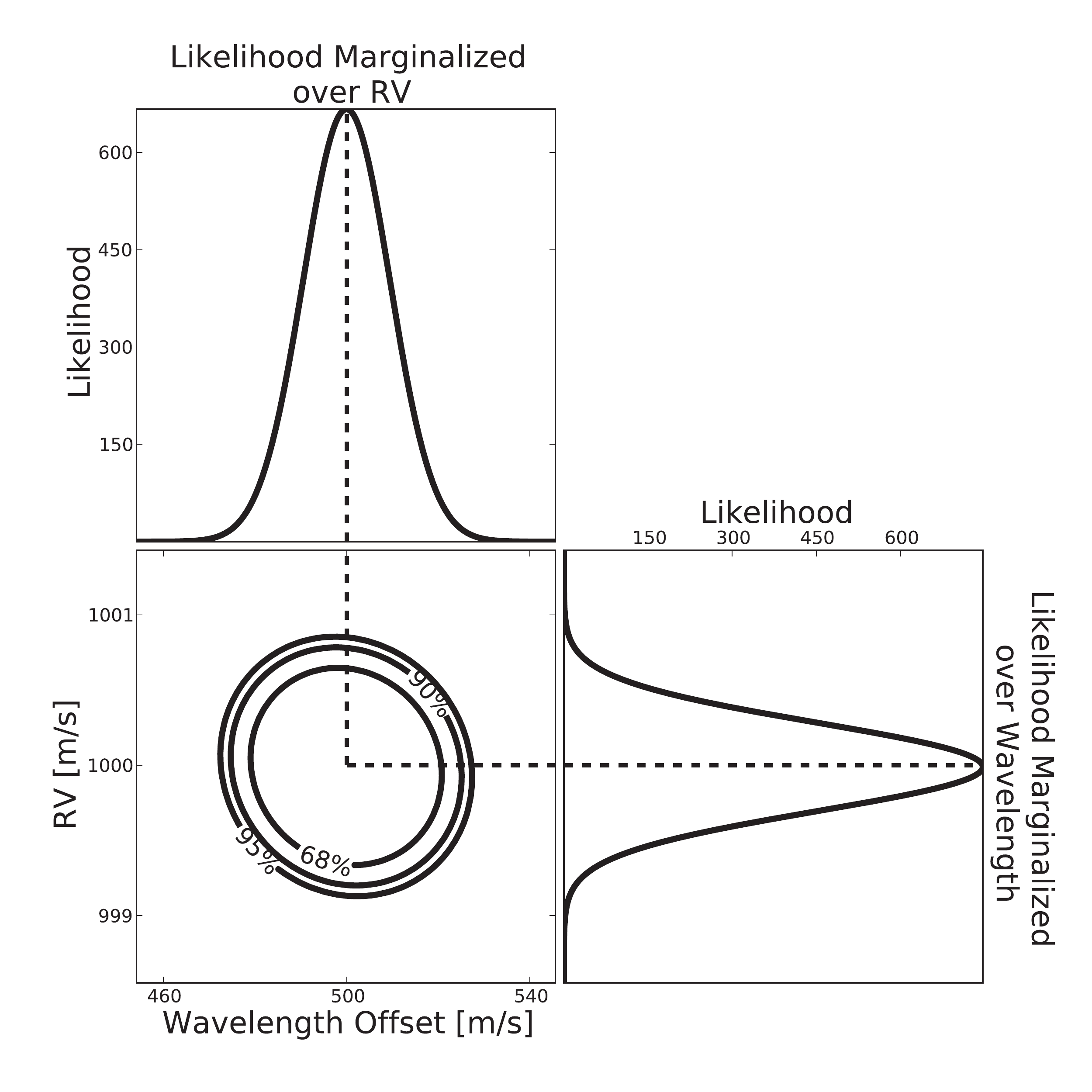}
      \caption{\small Simulated likelihood contours describing the radial velocity and wavelength calibration error reconstructions for injected values of 1000 m $\rm s^{-1}$ and 500 m $\rm s^{-1}$ respectively, for SNR=100 and a spectral bandwidth of $\Delta \lambda = 88\mbox{\AA}$. In contrast to traditional RV reconstruction methods, the contours show an elliptical shape, demonstrating that the RV and wavelength calibration error measurements are not highly correlated.}
    \label{fig:contours}
\end{figure}

In order to measure a radial velocity shift, two data sets, each represented by a figure like Figure \ref{fig:smile}, are produced: a template spectrum and an epoch spectrum with an RV shift and wavelength calibration error. The template spectrum is assumed to have infinite signal to noise, due to either a long on-sky integration or the use of a simulated spectrum. The template spectrum is multiplied by a complex phasor as in Equation \ref{eqn:compare_general}, containing a test RV shift, and is interpolated onto a new wavelength grid to represent the effect of the wavelength calibration error. This modified template spectrum and the originally shifted epoch spectrum are then compared using a chi-squared test that treats the real and imaginary parts of the complex visibilities separately:

\begin{equation}
\frac{-\chi^2}{2} = \displaystyle\sum\limits_{\nu} \left( B_{\nu}^{s} - B_{\nu}^{1} \right )^{2} + \left( C_{\nu}^{s} - C_{\nu}^{1} \right )^{2}
\label{eqn:loglike}
\end{equation}
where $B_{\nu}^{1}$ and $C_{\nu}^{1}$ describe the epoch spectrum, and $B_{\nu}^{s}$ and $C_{\nu}^{s}$ describe the modified template spectrum. Because the likelihood function is defined as
\begin{equation}
L = e^{\frac{-\chi^2}{2}},
\label{eqn:like}
\end{equation}
Equation \ref{eqn:loglike} is considered to be log($L$). We compute log($L$) on a grid of radial velocity and wavelength calibration error test points, and find the points that minimize Equation \ref{eqn:loglike} by parabola fitting. By computing many such solutions for independent realizations of the noise, we construct histograms of radial velocity and wavelength calibration errors solutions that are well described by Gaussian statistics. We define the standard deviation of 100 radial velocity solutions to be the ``radial velocity precision." By fitting a 2D Gaussian distribution to a 2D histogram of several thousand radial velocity and wavelength calibration error solutions, we compute error ellipses demonstrating the relationship between the two parameters (Figure \ref{fig:contours}). In contrast to traditional RV reconstruction methods, the contours show an elliptical shape, demonstrating that the RV and wavelength calibration error measurements are not highly correlated.

\begin{table*}[t]

\small
\centering
\caption{Simulation Parameters}
\label{tbl:assump1}
\begin{tabular}{ p{4.5cm} p{1.5cm} p{1.0cm} p{9.5cm}}
 & & & \\
\hline \hline
Parameter                     & Value     & Units      & Description \\
\hline 
Spectrograph resolution       & 10,000     &    & This is chosen to be small to avoid resolving the interferometer comb \\
Pixels per resolution element & 4.0         & pixels     & Slightly exceeds Nyquist sampling \\
Bandpass                      & 507-595   & nm         & $V$-band FWHM\\
Stellar model                 & Coelho    & N/A        & G star spectrum from Coelho et al 2007\\
Scan step size                & 0.012     & $\mu$m    & 20$\times$(Nyquist sampling) \\
Scan stroke                   & 2.5       & $\mu$m    & 5 fringe cycles \\
Bulk Delay                    & 2.0         & cm         & \\

\hline
\end{tabular}
\end{table*}

\begin{figure}[t]
  \centering
      \includegraphics[width=0.5\textwidth]{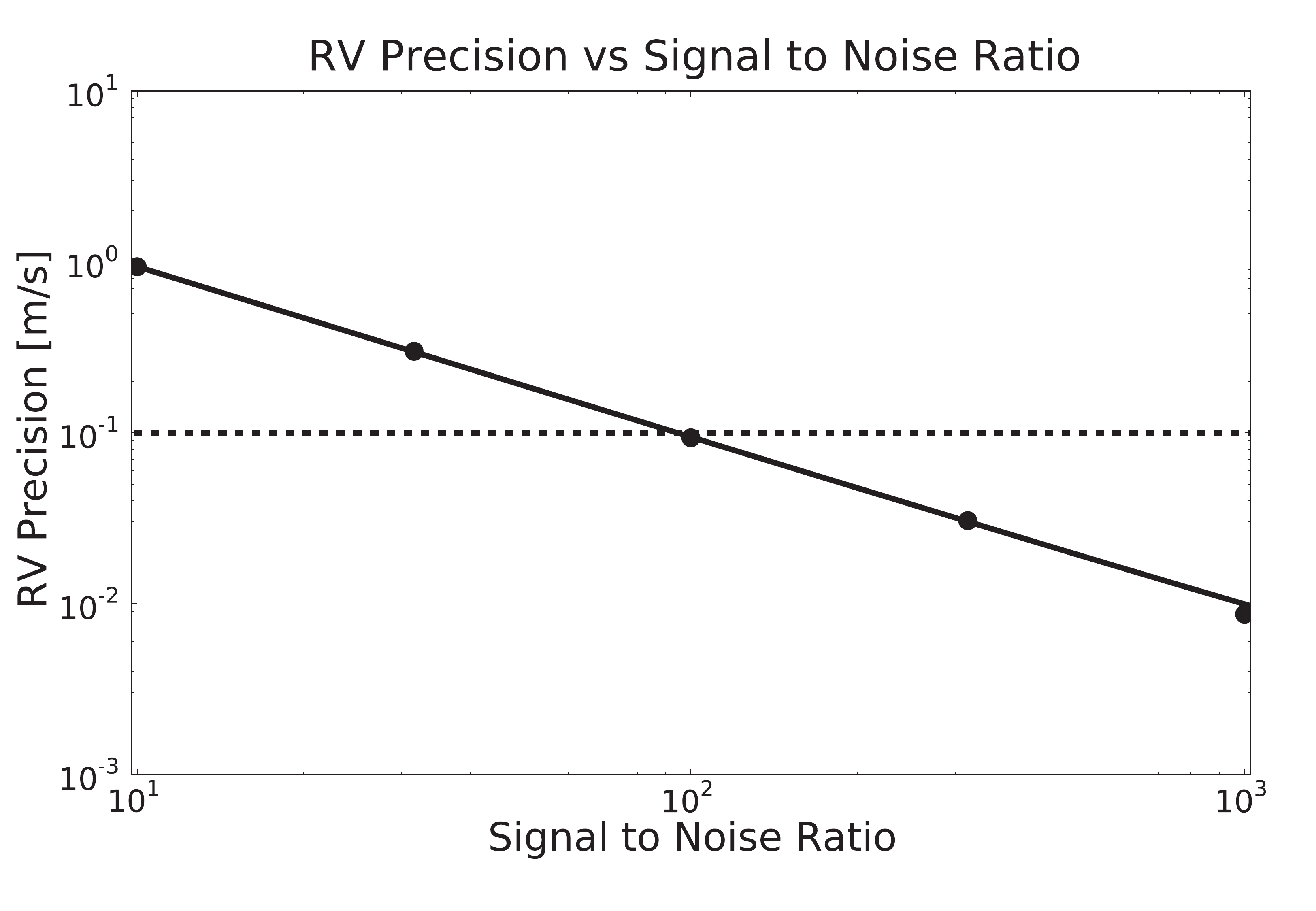}
      \caption{\small Radial velocity precision is plotted against the signal to noise ratio for Poisson noise limited measurements, (no instrumental or astrophysical noise sources). Under these conditions, an SNR of about 94 per phase step, for 200 phase steps, is required to reach a precision of 10 cm $\rm s^{-1}$. The RV precisions shown in this plot were calculated using a spectral bandwidth of $\Delta \lambda = 88\mbox{\AA}$, and were divided by $\sqrt{10}$ to reflect the RV precisions associated with the full $V$-band bandwidth of $880\mbox{\AA}$. }
    \label{fig:snr}
\end{figure}

\begin{figure}[t]
  \centering
      \includegraphics[width=0.5\textwidth]{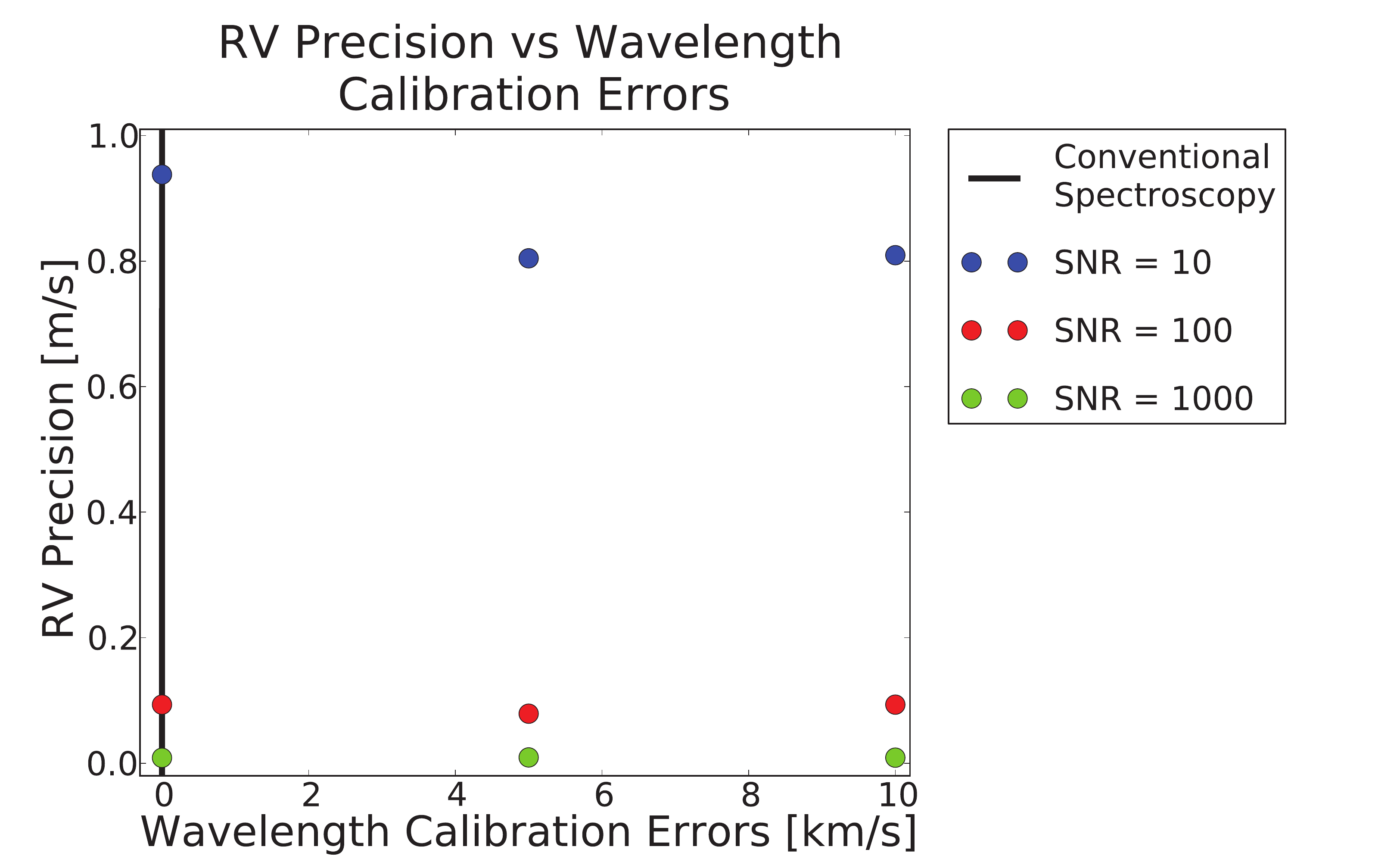}
      \caption{\small The radial velocity precision is shown to be approximately constant over 10 km $\rm s^{-1}$ ($\sim 0.2 \mbox{\AA}$) wavelength calibration errors for SNR = 10, 100, and 1000. In contrast, the radial velocity precision is proportional to the wavelength calibration error in conventional spectroscopy. The RV precisions shown in this plot were calculated using a spectral bandwidth of $\Delta \lambda = 88\mbox{\AA}$, and were divided by $\sqrt{10}$ to reflect the RV precisions associated with the full $V$-band bandwidth of $880\mbox{\AA}$. }
    \label{fig:wav}
\end{figure}

\begin{figure}[h!]
  \centering
      \includegraphics[width=0.5\textwidth]{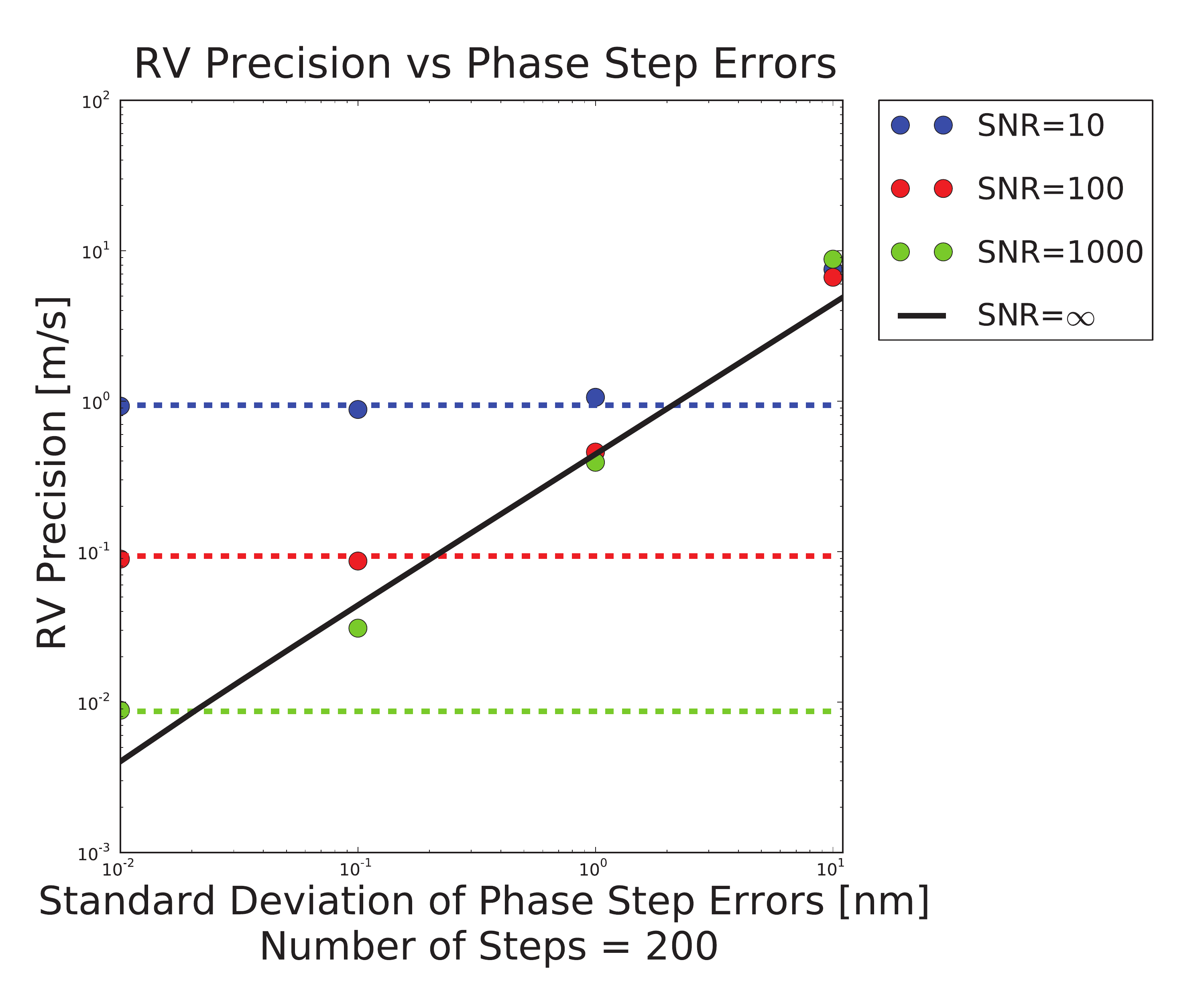}
      \caption{\small The RV precision is plotted against the standard deviation of the phase step error. The horizontal dotted lines represent the Poisson limited RV precision for each SNR. In the phase step error limited regions, the RV precisions approach the limiting $SNR=\infty$ condition, shown by the black line. For SNR = 100, $\Delta \tau$ must be known to less than 1 nm to reach the Poisson limited regime. Because the phase step errors are assumed to be uncorrelated, the total number of phase steps affects the final RV precision. In this simulation, we chose 200 steps (Table \ref{tbl:assump1}). The RV precisions shown in this plot were calculated using a spectral bandwidth of $\Delta \lambda = 88\mbox{\AA}$, and were divided by $\sqrt{10}$ to reflect the RV precisions associated with the full $V$-band bandwidth of $880\mbox{\AA}$. }
    \label{fig:phase_steps}
\end{figure}

\section{Error Budget}
\label{sec:sim_noise}

The simulation described in Section \ref{sec:sim} is used to analyze the affects of Poisson noise, wavelength calibration errors, and interferometer position errors. The assumptions used in the simulation are listed in Table \ref{tbl:assump1}. We also note that all radial velocity and wavelength calibration error solutions discussed in this section were generated using a spectral bandwidth of $\Delta \lambda = 88\mbox{\AA}$, while the FWHM of the $V$-band filter is $\Delta \lambda = 880\mbox{\AA}$. We chose this smaller bandwidth to accommodate the available computing resources. Because the number of spectral lines increases with the square root of the bandwidth, however, we can represent the $\Delta \lambda = 880\mbox{\AA}$ radial velocity precision by dividing the $\Delta \lambda = 88\mbox{\AA}$ radial velocity precision by $\sqrt{10}$. The text and figures below refer to radial velocity precisions that have been modified by this factor.

\subsection{Poisson Noise}

In the absence of all instrumental or astrophysical noise sources, the radial velocity precision decreases as $1/SNR$ (Figure \ref{fig:snr}). Under these conditions, a signal to noise ratio of about 94 per spectrograph resolution element is required to reach a radial velocity precision of 10 cm $\rm s^{-1}$. Sections \ref{sec:wav_cal}-\ref{sec:int_errors} below describe how instrumental noise sources cause the RV precision to deviate from this ideal case. 

\subsection{Wavelength Calibration Errors}
\label{sec:wav_cal}

Figure \ref{fig:wav} shows that the radial velocity precision changes by mm/s per km/s of wavelength calibration error for high signal to noise ratios.  The black line in Figure \ref{fig:wav} represents the 1:1 correspondence between wavelength calibration and radial velocity using conventional RV reconstruction methods. We have therefore demonstrated that by temporally varying a stellar spectrum, the dependence of the radial velocity precision on a consistent, correct wavelength solution is reduced. 

\subsection{Interferometer Position Errors}
\label{sec:int_errors}

This robustness against wavelength calibration errors comes at the cost of precise knowledge of the interferometer position. The interferometer position is described by a constant ``bulk" offset and a much smaller time varying phase shift ($\tau_{0}$ and $\Delta \tau$, respectively, in Equation \ref{eqn:decomposed_intensity}).

Equation \ref{eqn:compare_general} shows that errors in estimation of the bulk delay produce proportional errors in the RV precision. Therefore, the same error in a bulk delay measurement will result in poorer RV precision for stars with larger radial velocities. In order to calculate the smallest required bulk delay measurement error, we must consider the target stars with the largest radial velocities. Barycentric motion produces radial velocity differences of up to 60 km $\rm s^{-1}$. In order to reach an RV precision of 10 cm $\rm s^{-1}$ with a typical bulk delay of $2\,$cm, the maximum bulk delay measurement error is $\left ( \frac{10 \, \mbox{cm}}{60 \, \mbox{km}}\right )2 \, \mbox{cm} = 33 \,$nm. State-of-the-art piezo electric stages are capable of sub-nm RMS measurement accuracy (e.g. ~\citet{samuele_experimental_2007}), so we conclude that the bulk delay can be adequately measured to provide 10 cm s$^{-1}$  precision on the stars with the largest radial velocities.

\begin{table*}[b]

\small
\centering
\caption{Instrument-Specific Parameters}
\label{tbl:extra_assumps}
\begin{tabular}{ p{6.0cm} p{1.1cm} p{0.75cm} p{8.0cm}}
 & & & \\
\hline \hline
Parameter        & Value     & Units     & Comments \\
\hline 
Bandpass    & 507-595   & nm    & $V$-band FWHM\\
Sky to detector throughput & 7.4   & $\%$   & Representative of a single mode fiber coupled to a $30\%$ Strehl input beam and a spectrograph with $30\%$ efficiency\\
Interferometer position control requirement  & 1  & nm   & Conservative requirement based on 0.4nm state-of-the-art piezoelectric parameters\\
Stellar visual magnitude  & 8.5     & $m_{v}$ & Representative of the mode of the visual magnitudes of the 100 brightest G stars\\
Minimum SNR per resolution element & 8 &  & Based on minimum SNR to avoid the read noise limited regime\\
\hline
\end{tabular}
\end{table*}


Errors in the phase step estimations, however, impose more strict metrology requirements. Figure \ref{fig:phase_steps} shows the radial velocity precision as a function of phase step error. Normally distributed errors with standard deviations represented by the values on the x-axis were added to the true phase delays. The horizontal regions of the SNR = 10, 100, and 1000 plots show the Poisson limited regime, while the sloped regions represent the phase step error limited regime. The black line represents SNR = $\infty$. For a single SNR = 100 scan, the phase steps must to known to sub-nanometer precision to remain in the Poisson limited regime. Because the phase step  position errors are expected to be normally distributed, however, taking multiple scans will improve the RV precision by the square root of the number of scans. In this way, the SNR of the combined scan is increased while the errors due to interferometer position errors are reduced. Section \ref{sec:sim_results} describes this approach in detail.


\section{The Feasibility of an Earth-Sun Analog Measurement}
\label{sec:sim_results}

We have shown that the modulation of a stellar spectrum in time decouples wavelength calibration errors from RV measurements, while coupling interferometer position errors to RV measurements. Given these new constraints, we now address the feasibility of reaching a radial velocity precision of 8.9 cm $\rm s^{-1}$ on a G-type star using an exiting telescope. Our assumed bandpass, throughput, interferometer control requirement, and target star magnitude are summarized in Table \ref{tbl:extra_assumps} and described in detail below. 

Operating in the visible ($\approx 500-600\,$nm) while observing Sun-like stars has the dual advantages of covering the peak of the G star blackbody function while reducing contamination due to telluric lines compared to the infrared. A visible bandpass, however, may come at the cost of throughput. In order to minimize the effects of interferometer position errors, calibration and star light should be common path in the interferometer and spectrograph. One possible approach would be to feed the instrument with a single mode fiber, containing both starlight and light from a calibration source (for example, a frequency stabilized laser). In this way, modulation of the calibration light, imprinted on the science data itself, will measure the changes in the optical path difference of the interferometer. 

The single-mode fiber coupling efficiency is determined primarily by the Strehl ratio of the associated adaptive optics system. Single-mode fibers are inherently diffraction limited, and require diffraction-limited beams delivered from a telescope in order to have any consequential coupling: seeing-limited telescopes will not couple nearly enough starlight for a Doppler survey of nearby stars ~\citep{shaklan_coupling_1988}.  Recently, however, advanced adaptive optics systems have reported significant Strehl ratios at visible wavelengths.  For example, the Robo-AO system on the Palomar 60-inch telescope has achieved a Strehl ratio of $18\%$ in i-band ~\citep{baranec_robo-ao:_2012}.  The Palm-3000 adaptive optics system on the Palomar 200-inch telescope is designed to achieve high Strehl ratios at visible wavelengths. Palm-3000 is predicted to achieve $30\%$ Strehl in V-band, equivalent to a 95 nm RMS wavefront error ~\citep{dekany_palm-3000:_2013}.  Researchers at the University of Arizona recently commissioned a visible-light adaptive optics system (VisAO) on the 6.5-m Magellan Clay Telescope, demonstrating a $43\%$ Strehl ratio, or 149 nm RMS wavefront error, in Y-band ~\citep{close_diffraction-limited_2013}. For the purposes of this study, we estimate the total sky-to-detector throughput to be $7.4\%$ by multiplying the ideal fiber coupling efficiency of $82\%$ by a Strehl ratio of $30\%$, based on the performance of Palm-3000 at visible wavelengths, and a spectrograph efficiency of $30\%$. 

As described in Section \ref{sec:theory}, instrumental noise varying more slowly than the interferometer's phase steps is rejected. For maximum noise rejection, the interferometer should therefore scan as quickly as possible. The interferometer's speed is then limited by detector frame rates and read noise. The lowest noise detectors at optical wavelengths, CCDs, have significant readout times of seconds to minutes. This limits the rate at which we can modulate an optical signal to the point that lock-in amplification is effectively pointless with most CCD detectors. However, the development of high-frame rate, low read noise EMCCDs and sCMOS detectors opens up the possibility of using lock-in amplification at optical wavelengths. Assuming a detector read noise of $2 e^{-}$ per pixel, gain  = 1, and four pixels per resolution element, the detector must measure 64 photons per resolution element (SNR=8) to remain well within the photon noise limited regime. The exposure time necessary to reach this minimum SNR will place an upper limit on the scanning speed. 

Taking a series of scans can reduce normally distributed noise due to interferometer position errors. The number of scans is chosen such that the desired RV precision is achieved under realistic position error conditions. State-of-the-art piezoelectric stages can be controlled to less than 1 nm RMS (e.g. ~\citet{samuele_experimental_2007} controlled the Physik Instrumente P-752 flexure stage to 0.4 nm rms), so we choose a 1 nm control requirement. In order to reach 10 cm $\rm s^{-1}$ with phase step errors of 1 nm RMS and bulk delay errors of $10\%$ $\tau_{0}$ , we require 139 scans, each with SNR = 8. 

We now choose a target star magnitude and telescope size to determine the observing time required to take 139 such scans. To choose a representative target star magnitude, we constructed a histogram of the 100 brightest G stars. We choose $m_{v} = 8.5$ as a representative magnitude from this sample. 

The Palomar 200" telescope would take about 0.074 seconds to reach SNR=8 per resolution element, given the assumptions in Tables \ref{tbl:assump1} and \ref{tbl:extra_assumps}. For 139 scans, each with 200 $0.074\,$s phase steps, the 200" telescope would require about 0.57 hours to reach a radial velocity precision of 10 cm $\rm s^{-1}$. It is therefore possible to reach the radial velocity precision necessary to detect Earth-like planets around Sun-like stars using existing, 5-meter-class telescopes.

\section{Conclusions}
\label{sec:conclusion}

The current state-of-the-art radial velocity instruments are limited by their ability to maintain their system's wavelength solution in the presence of slowly varying instrumental noise. Lock-in amplification, however, can suppress such long timescale noise sources, while decoupling the effects of wavelength calibration errors from radial velocity precision. The simulations presented in this paper indicate that lock-in amplified, externally-dispersed interferometry is a possible path forward to reach the radial velocity precision necessary to detect Earth-Sun analog systems on existing, medium-sized telescopes.  

\acknowledgements

This material is based upon work supported by the National Science Foundation Graduate Research Fellowship under Grant No. DGE-1144469. P.S.M. acknowledges support for his work from the Hubble Fellowship Program, provided by NASA through Hubble Fellowship grant HST-HF-51326.01-A awarded by the STScI, which is operated by the AURA, Inc., for NASA, under contract NAS 5-26555. P.S.M, G.V., and J.K.W. were supported by the Directors Research and Development Fund and Caltech-JPL President and Director's fund at the California Institute of Technology/Jet Propulsion Laboratory. M.B. is supported by a National Space Technology Research Fellowship. J.A.J. is supported by generous grants from the David and Lucile Packard Foundation and the Alfred P. Sloan Foundation. JKW and GV are supported by the We would like to thank Lynne Hillenbrand for useful discussions.

\bibliography{ms}

\begin{thebibliography}{}
\expandafter\ifx\csname natexlab\endcsname\relax\def\natexlab#1{#1}\fi

\bibitem[{Baranec {et~al.}(2012)Baranec, Riddle, Ramaprakash, Law, Tendulkar,
  Kulkarni, Dekany, Bui, Davis, Burse, Das, Hildebrandt, Punnadi, \&
  Smith}]{baranec_robo-ao:_2012}
Baranec, C., Riddle, R., Ramaprakash, A.~N., {et~al.} 2012, {arXiv:1210.0532}
  [astro-ph], 844704, proc. {SPIE} 8447, Adaptive Optics Systems {III}, 844704
  (September 13, 2012)

\bibitem[{Baranec {et~al.}(2013)Baranec, Riddle, Law, Ramaprakash, Tendulkar,
  Hogstrom, Bui, Burse, Chordia, Das, \& Dekany}]{baranec_rise_2013}
Baranec, C., Riddle, R., Law, N., {et~al.} 2013, Proceedings of the Advanced
  Maui Optical and Space Surveillance Technologies Conference, held in Wailea,
  Maui, Hawaii, September 10-13, 2013, Ed.: S. Ryan, The Maui Economic
  Development Board, E48

\bibitem[{Baranne {et~al.}(1996)Baranne, Queloz, Mayor, Adrianzyk, Knispel,
  Kohler, Lacroix, Meunier, Rimbaud, \& Vin}]{baranne_elodie:_1996}
Baranne, A., Queloz, D., Mayor, M., {et~al.} 1996, Astronomy and Astrophysics
  Supplement Series, 119, 373

\bibitem[{{Batalha} {et~al.}(2013){Batalha}, {Rowe}, {Bryson}, {Barclay},
  {Burke}, {Caldwell}, {Christiansen}, {Mullally}, {Thompson}, {Brown},
  {Dupree}, {Fabrycky}, {Ford}, {Fortney}, {Gilliland}, {Isaacson}, {Latham},
  {Marcy}, {Quinn}, {Ragozzine}, {Shporer}, {Borucki}, {Ciardi}, {Gautier},
  {Haas}, {Jenkins}, {Koch}, {Lissauer}, {Rapin}, {Basri}, {Boss}, {Buchhave},
  {Carter}, {Charbonneau}, {Christensen-Dalsgaard}, {Clarke}, {Cochran},
  {Demory}, {Desert}, {Devore}, {Doyle}, {Esquerdo}, {Everett}, {Fressin},
  {Geary}, {Girouard}, {Gould}, {Hall}, {Holman}, {Howard}, {Howell},
  {Ibrahim}, {Kinemuchi}, {Kjeldsen}, {Klaus}, {Li}, {Lucas}, {Meibom},
  {Morris}, {Pr{\v s}a}, {Quintana}, {Sanderfer}, {Sasselov}, {Seader},
  {Smith}, {Steffen}, {Still}, {Stumpe}, {Tarter}, {Tenenbaum}, {Torres},
  {Twicken}, {Uddin}, {Van Cleve}, {Walkowicz}, \& {Welsh}}]{Batalha2013}
{Batalha}, N.~M., {Rowe}, J.~F., {Bryson}, S.~T., {et~al.} 2013, \apjs, 204, 24

\bibitem[{Berta {et~al.}(2013)Berta, Irwin, \&
  Charbonneau}]{berta_constraints_2013}
Berta, Z.~K., Irwin, J., \& Charbonneau, D. 2013, The Astrophysical Journal,
  775, 91

\bibitem[{{Borucki} {et~al.}(2011){Borucki}, {Koch}, {Basri}, {Batalha},
  {Brown}, {Bryson}, {Caldwell}, {Christensen-Dalsgaard}, {Cochran}, {DeVore},
  {Dunham}, {Gautier}, {Geary}, {Gilliland}, {Gould}, {Howell}, {Jenkins},
  {Latham}, {Lissauer}, {Marcy}, {Rowe}, {Sasselov}, {Boss}, {Charbonneau},
  {Ciardi}, {Doyle}, {Dupree}, {Ford}, {Fortney}, {Holman}, {Seager},
  {Steffen}, {Tarter}, {Welsh}, {Allen}, {Buchhave}, {Christiansen}, {Clarke},
  {Das}, {D{\'e}sert}, {Endl}, {Fabrycky}, {Fressin}, {Haas}, {Horch},
  {Howard}, {Isaacson}, {Kjeldsen}, {Kolodziejczak}, {Kulesa}, {Li}, {Lucas},
  {Machalek}, {McCarthy}, {MacQueen}, {Meibom}, {Miquel}, {Prsa}, {Quinn},
  {Quintana}, {Ragozzine}, {Sherry}, {Shporer}, {Tenenbaum}, {Torres},
  {Twicken}, {Van Cleve}, {Walkowicz}, {Witteborn}, \& {Still}}]{Borucki2011}
{Borucki}, W.~J., {Koch}, D.~G., {Basri}, G., {et~al.} 2011, \apj, 736, 19

\bibitem[{Close {et~al.}(2013)Close, Males, Morzinski, Kopon, Follette,
  Rodigas, Hinz, Wu, Puglisi, Esposito, Riccardi, Pinna, Xompero, Briguglio,
  Uomoto, \& Hare}]{close_diffraction-limited_2013}
Close, L.~M., Males, J.~R., Morzinski, K., {et~al.} 2013, The Astrophysical
  Journal, 774, 94

\bibitem[{Coelho {et~al.}(2007)Coelho, Bruzual, Charlot, Weiss, Barbuy, \&
  Ferguson}]{coelho_spectral_2007}
Coelho, P., Bruzual, G., Charlot, S., {et~al.} 2007, Monthly Notices of the
  Royal Astronomical Society, 382, 498

\bibitem[{Dekany {et~al.}(2013)Dekany, Roberts, Burruss, Bouchez, Truong,
  Baranec, Guiwits, Hale, Angione, Trinh, Zolkower, Shelton, Palmer, Henning,
  Croner, Troy, {McKenna}, Tesch, Hildebrandt, \&
  Milburn}]{dekany_palm-3000:_2013}
Dekany, R., Roberts, J., Burruss, R., {et~al.} 2013, The Astrophysical Journal,
  776, 130

\bibitem[{{Deming} {et~al.}(2009){Deming}, {Seager}, {Winn}, {Miller-Ricci},
  {Clampin}, {Lindler}, {Greene}, {Charbonneau}, {Laughlin}, {Ricker},
  {Latham}, \& {Ennico}}]{Deming2009}
{Deming}, D., {Seager}, S., {Winn}, J., {et~al.} 2009, \pasp, 121, 952

\bibitem[{{Dhillon} \& {Marsh}(2001)}]{Dhillon2001}
{Dhillon}, V., \& {Marsh}, T. 2001, \nar, 45, 91

\bibitem[{Dressing \& Charbonneau(2013)}]{dressing_occurrence_2013}
Dressing, C.~D., \& Charbonneau, D. 2013, The Astrophysical Journal, 767, 95

\bibitem[{{Dumusque} {et~al.}(2011){Dumusque}, {Udry}, {Lovis}, {Santos}, \&
  {Monteiro}}]{Dumusque2011}
{Dumusque}, X., {Udry}, S., {Lovis}, C., {Santos}, N.~C., \& {Monteiro},
  M.~J.~P.~F.~G. 2011, \aap, 525, A140

\bibitem[{Dumusque {et~al.}(2012)Dumusque, Pepe, Lovis, SŽgransan, Sahlmann,
  Benz, Bouchy, Mayor, Queloz, Santos, \& Udry}]{dumusque_earth-mass_2012}
Dumusque, X., Pepe, F., Lovis, C., {et~al.} 2012, Nature, 491, 207

\bibitem[{{Erskine}(2003)}]{Erskine2003}
{Erskine}, D.~J. 2003, \pasp, 115, 255

\bibitem[{{Erskine} {et~al.}(2003){Erskine}, {Edelstein}, {Feuerstein}, \&
  {Welsh}}]{Erskine2003b}
{Erskine}, D.~J., {Edelstein}, J., {Feuerstein}, W.~M., \& {Welsh}, B. 2003,
  \apjl, 592, L103

\bibitem[{{Ge} {et~al.}(2006){Ge}, {van Eyken}, {Mahadevan}, {DeWitt}, {Kane},
  {Cohen}, {Vanden Heuvel}, {Fleming}, {Guo}, {Henry}, {Schneider}, {Ramsey},
  {Wittenmyer}, {Endl}, {Cochran}, {Ford}, {Mart{\'{\i}}n}, {Israelian},
  {Valenti}, \& {Montes}}]{Ge2006}
{Ge}, J., {van Eyken}, J., {Mahadevan}, S., {et~al.} 2006, \apj, 648, 683

\bibitem[{{Hajian} {et~al.}(2007){Hajian}, {Behr}, {Cenko}, {Olling},
  {Mozurkewich}, {Armstrong}, {Pohl}, {Petrossian}, {Knuth}, {Hindsley},
  {Murison}, {Efroimsky}, {Dantowitz}, {Kozubal}, {Currie}, {Nordgren},
  {Tycner}, \& {McMillan}}]{Hajian2007}
{Hajian}, A.~R., {Behr}, B.~B., {Cenko}, A.~T., {et~al.} 2007, \apj, 661, 616

\bibitem[{{Howard} {et~al.}(2010){Howard}, {Marcy}, {Johnson}, {Fischer},
  {Wright}, {Isaacson}, {Valenti}, {Anderson}, {Lin}, \& {Ida}}]{Howard2010}
{Howard}, A.~W., {Marcy}, G.~W., {Johnson}, J.~A., {et~al.} 2010, Science, 330,
  653

\bibitem[{Howard {et~al.}(2011)Howard, Johnson, Marcy, Fischer, Wright, Henry,
  Isaacson, Valenti, Anderson, \& Piskunov}]{howard_nasa-uc_2011}
Howard, A.~W., Johnson, J.~A., Marcy, G.~W., {et~al.} 2011, The Astrophysical
  Journal, 726, 73

\bibitem[{Howard {et~al.}(2012)Howard, Marcy, Bryson, Jenkins, Rowe, Batalha,
  Borucki, Koch, Dunham, Gautier, Van~Cleve, Cochran, Latham, Lissauer, Torres,
  Brown, Gilliland, Buchhave, Caldwell, Christensen-Dalsgaard, Ciardi, Fressin,
  Haas, Howell, Kjeldsen, Seager, Rogers, Sasselov, Steffen, Basri,
  Charbonneau, Christiansen, Clarke, Dupree, Fabrycky, Fischer, Ford, Fortney,
  Tarter, Girouard, Holman, Johnson, Klaus, Machalek, Moorhead, Morehead,
  Ragozzine, Tenenbaum, Twicken, Quinn, Isaacson, Shporer, Lucas, Walkowicz,
  Welsh, Boss, Devore, Gould, Smith, Morris, Prsa, Morton, Still, Thompson,
  Mullally, Endl, \& {MacQueen}}]{howard_planet_2012}
Howard, A.~W., Marcy, G.~W., Bryson, S.~T., {et~al.} 2012, The Astrophysical
  Journal Supplement Series, 201, 15

\bibitem[{{Koch} {et~al.}(2010){Koch}, {Borucki}, {Basri}, {Batalha}, {Brown},
  {Caldwell}, {Christensen-Dalsgaard}, {Cochran}, {DeVore}, {Dunham},
  {Gautier}, {Geary}, {Gilliland}, {Gould}, {Jenkins}, {Kondo}, {Latham},
  {Lissauer}, {Marcy}, {Monet}, {Sasselov}, {Boss}, {Brownlee}, {Caldwell},
  {Dupree}, {Howell}, {Kjeldsen}, {Meibom}, {Morrison}, {Owen}, {Reitsema},
  {Tarter}, {Bryson}, {Dotson}, {Gazis}, {Haas}, {Kolodziejczak}, {Rowe}, {Van
  Cleve}, {Allen}, {Chandrasekaran}, {Clarke}, {Li}, {Quintana}, {Tenenbaum},
  {Twicken}, \& {Wu}}]{Koch2010}
{Koch}, D.~G., {Borucki}, W.~J., {Basri}, G., {et~al.} 2010, \apjl, 713, L79

\bibitem[{Lovis \& Pepe(2007)}]{lovis_new_2007}
Lovis, C., \& Pepe, F. 2007, Astronomy and Astrophysics, 468, 1115

\bibitem[{Mahadevan {et~al.}(2008)Mahadevan, van Eyken, Ge, {DeWitt}, Fleming,
  Cohen, Crepp, \& Vanden~Heuvel}]{mahadevan_measuring_2008}
Mahadevan, S., van Eyken, J., Ge, J., {et~al.} 2008, The Astrophysical Journal,
  678, 1505

\bibitem[{Muirhead {et~al.}(2011)Muirhead, Edelstein, Erskine, Wright,
  Muterspaugh, Covey, Wishnow, Hamren, Andelson, Kimber, Mercer, Halverson,
  Vanderburg, Mondo, Czeszumska, \& Lloyd}]{muirhead_precise_2011}
Muirhead, P.~S., Edelstein, J., Erskine, D.~J., {et~al.} 2011, Publications of
  the Astronomical Society of the Pacific, 123, 709, {ArticleType:}
  research-article / Full publication date: June 2011 / Copyright © 2011 The
  University of Chicago Press

\bibitem[{Murphy {et~al.}(2007)Murphy, Udem, Holzwarth, Sizmann, Pasquini,
  Araujo-Hauck, Dekker, {D'Odorico}, Fischer, HŠnsch, \&
  Manescau}]{murphy_high-precision_2007}
Murphy, M.~T., Udem, T., Holzwarth, R., {et~al.} 2007, Monthly Notices of the
  Royal Astronomical Society, 380, 839, {arXiv:astro-ph/0703622}

\bibitem[{Pepe {et~al.}(2011)Pepe, Lovis, Segransan, Benz, Bouchy, Dumusque,
  Mayor, Queloz, Santos, \& Udry}]{pepe_harps_2011}
Pepe, F., Lovis, C., Segransan, D., {et~al.} 2011, Astronomy \& Astrophysics,
  534, A58

\bibitem[{Petigura {et~al.}(2013)Petigura, Howard, \& Marcy}]{Petigura2013}
Petigura, E.~A., Howard, A.~W., \& Marcy, G.~W. 2013, Proceedings of the
  National Academy of Sciences, 110, 19273

\bibitem[{Ricker(2014)}]{ricker_transiting_2014}
Ricker, G.~R. 2014, Journal of the American Association of Variable Star
  Observers ({JAAVSO}), 42, 234

\bibitem[{Samuele {et~al.}(2007)Samuele, Wallace, Schmidtlin, Shao, Levine, \&
  Fregoso}]{samuele_experimental_2007}
Samuele, R., Wallace, J., Schmidtlin, E., {et~al.} 2007, in 2007 {IEEE}
  Aerospace Conference, 1--7

\bibitem[{Shaklan \& Roddier(1988)}]{shaklan_coupling_1988}
Shaklan, S., \& Roddier, F. 1988, Applied Optics, 27, 2334

\bibitem[{{Steinmetz} {et~al.}(2008){Steinmetz}, {Wilken}, {Araujo-Hauck},
  {Holzwarth}, {H{\"a}nsch}, {Pasquini}, {Manescau}, {D'Odorico}, {Murphy},
  {Kentischer}, {Schmidt}, \& {Udem}}]{Steinmetz2008}
{Steinmetz}, T., {Wilken}, T., {Araujo-Hauck}, C., {et~al.} 2008, Science, 321,
  1335

\bibitem[{{van Eyken} {et~al.}(2010){van Eyken}, {Ge}, \&
  {Mahadevan}}]{Vaneyken2010}
{van Eyken}, J.~C., {Ge}, J., \& {Mahadevan}, S. 2010, \apjs, 189, 156

\bibitem[{Wilken {et~al.}(2012)Wilken, Curto, Probst, Steinmetz, Manescau,
  Pasquini, Gonz‡lez~Hern‡ndez, Rebolo, HŠnsch, Udem, \&
  Holzwarth}]{wilken_spectrograph_2012}
Wilken, T., Curto, G.~L., Probst, R.~A., {et~al.} 2012, Nature, 485, 611

\end{thebibliography}
\bibliographystyle{apj}

\setcounter{figure}{0}

%


\end{document}